%% file: chiir26-59.tex
\documentclass[sigconf,natbib=true]{acmart}

\usepackage{multirow}
\usepackage{graphicx} 
\usepackage{float} 
\usepackage{subfigure} 
\usepackage{{booktabs}}
\usepackage{tikz}
\usepackage{tikz-cd}
\usepackage{pgfplots}
\usepackage{adjustbox}
\usepackage{setspace}
\usepackage{tabularx}
\usepackage{framed}
\usepackage{enumitem}
\usepackage{mathrsfs} 
\usepackage{tcolorbox}
\usepackage{booktabs}
\usepackage{longtable}
\usepackage{subcaption}
\usepackage{siunitx}
\usepackage{multirow}
\usepackage[table]{xcolor}
\renewcommand\arraystretch{1.6}

\newcommand{\paratitle}[1]{\vspace{1.0ex}\noindent\textbf{#1}}

\renewcommand{\arraystretch}{0.9} % default is 1.0

\newtcolorbox{promptbox}{
  colback=gray!10,
  colframe=gray!80,
  boxrule=0.5pt,
  arc=4pt,
  left=6pt,
  right=6pt,
  top=6pt,
  bottom=6pt,
  fontupper=\itshape,
  title= Prompt
}

\pgfplotsset{compat=1.14}
\AtBeginDocument{%
  }

\copyrightyear{2026}
\acmYear{2026}
\setcopyright{cc}
\setcctype{by}
\acmConference[CHIIR '26]{2026 ACM SIGIR Conference on Human Information Interaction and Retrieval}{March 22--26, 2026}{Seattle, WA, USA}
\acmBooktitle{2026 ACM SIGIR Conference on Human Information Interaction and Retrieval (CHIIR '26), March 22--26, 2026, Seattle, WA, USA}
\acmPrice{}
\acmDOI{10.1145/3786304.3787942}
\acmISBN{979-8-4007-2414-5/2026/03}

\begin{document}

\title[How SERPs and AI-generated Podcasts Interact to Influence User Attitudes on Controversial Topics]{From \textit{SERPs} to Sound: How Search Engine Result Pages and AI-generated Podcasts Interact to Influence User Attitudes on Controversial Topics}


\author{Junjie Wang}
\affiliation{%
 \institution{Delft University of Technology}
 \city{Delft}
 \state{South Holland}
 \country{Netherlands}}
\email{j.wang-101@student.tudelft.nl}

\author{Gaole He}
\affiliation{%
\institution{Delft University of Technology}
 \city{Delft}
 \state{South Holland}
 \country{Netherlands}}
\email{g.he@tudelft.nl}

\author{Alisa Rieger}
\affiliation{%
\institution{GESIS, Leibniz Institute of the Social Sciences}
 \city{Cologne}
 \state{}
 \country{Germany}}
\email{alisa.rieger@gesis.org}

\author{Ujwal Gadiraju}\authornote{Corresponding author.}
\affiliation{%
\institution{Delft University of Technology}
 \city{Delft}
 \state{South Holland}
 \country{Netherlands}}
\email{u.k.gadiraju@tudelft.nl}


\newcommand{\nb}[3]{
  \fcolorbox{black}{#2}{\bfseries\sffamily\scriptsize#1}
    {\sf\small$\blacktriangleright$\textit{\textcolor{blue}{#3}}$\blacktriangleleft$}
}

\newcommand\ujwal[1]{\nb{Ujwal}{green}{#1}}

\begin{abstract} Compared to search engine result pages (SERPs), AI-generated podcasts represent a relatively new and relatively more passive modality of information consumption, delivering narratives in a naturally engaging format. As these two media increasingly converge in everyday information-seeking behavior, it is essential to explore how their interaction influences user attitudes, particularly in contexts involving controversial, value-laden, and often debated topics. Addressing this need, we aim to understand how information mediums of present-day SERPs and AI-generated podcasts interact to shape the opinions of users. To this end, through a controlled user study ($N=483$), we investigated user attitudinal effects of consuming information via SERPs and AI-generated podcasts, focusing on how the sequence and modality of exposure shape user opinions. A majority of users in our study corresponded to attitude change outcomes, and we found an effect of sequence on attitude change. Our results further revealed a role of viewpoint bias and the degree of topic controversiality in shaping attitude change, although we found no effect of individual moderators. 
\end{abstract}




\maketitle

\input{1_Introduction.tex}
\input{2_RelatedWork.tex}

\input{3_Method.tex}

\input{4_Results_new.tex}
\input{5_Discussion.tex}
\input{6_Conclusions.tex}

\section*{Acknowledgements}
%\begin{small}
We thank all the anonymous participants in our study. This work was supported by the TU Delft AI Initiative, the Model Driven Decisions Lab (\textit{MoDDL}), the \textit{ProtectMe} Convergence Flagship, and a grant from the TU Delft Fast Fund. We used Microsoft Copilot to assist with utility code for analyses and plots, and all generated code was independently reviewed and validated by the authors.    
%\end{small}

\bibliographystyle{ACM-Reference-Format}
\bibliography{9_references}
%\input{7_Appendix}


\end{document}

%% file: 1_Introduction.tex
\section{Introduction}
\label{sec:intro}

Traditional search engine result pages (SERPs) have served as the primary gateway to online information for years, offering users a curated list of sources that they can actively navigate, shaping user opinions and influencing their attitudes on a variety of topics~\cite{xu2021user,epstein2024search,epstein2017suppressing,white2015belief}. For example, there has been a wealth of evidence demonstrating how rankings of search results and their corresponding viewpoints influence consumer choices---called the \textit{search engine manipulation effect} (SEME)---since users tend to consume, trust, and rely more on higher-ranked web search results than lower-ranked results~\cite{draws2021not,epstein2024search}.
Over the last decade, we have witnessed a growing interest in the conversational search paradigm~\cite{radlinski2017theoretical} and a flourishing presence of digital assistants, chatbots~\cite{grudin2019chatbots}, and conversational agents~\cite{clark2019makes,jung2022great} that aid users in satisfying their information needs~\cite{avula2018searchbots,jung2019tell,liao2020conversational,avula2019embedding,trippas2020towards}. At the same time, there has been a growth in podcast consumption as a source of information~\cite{tobin2022people} with millions of episodes readily available on several topics, and the number of podcast listeners around the world has nearly doubled since 2019~\cite{statista2024podcasts,statista2024listeners,spotify2023podcastday}. Notably, podcasts uniquely deliver linear, narrative content that users typically consume with little interaction or selective control.

With advances in generative AI and the frantic adoption of AI-powered solutions in search systems over the last few years, there has been a shift in information seeking and consumption behaviors~\cite{white2025information}. For example, users today are widely exposed to AI-generated content in different media, such as textual \textit{AI overviews} on Google search~\cite{yang2025search+} or AI-generated podcasts in the audio medium using technologies like NotebookLM.\footnote{\url{https://notebooklm.google.com}}  Since digital information is now omnipresent and increasingly personalized, understanding how users form and update their opinions based on different media formats is critical. In recent work, \citet{white2025panmodal} argued that in the future, \textit{panmodality} (i.e., the use of multiple modalities either separately or collectively) will play a central role in information interaction. \citet{mayerhofer_blending_2025} explored how users engaged with an interface combining web search and a generative AI chat feature to solve health-related information tasks. Their qualitative findings suggest that while such integration with generative AI has the potential to enhance information-seeking, it can also lead to misplaced trust in favour of `ease-of-use' and seemingly perfect answers. This corroborates findings across the HCI and NLP research communities that has explored how users tend to build trust and reliance on agents powered by %large language models (
LLMs that demonstrably provide convincingly wrong answers or advice~\cite{si2024large,he2025plan,he2025conversational,biswas2025mind}.

In contrast to traditional SERPs, AI-generated podcasts represent a relatively new and more passive modality of information consumption, delivering synthesized narratives in a conversational format. 
Compared to other passive formats (e.g., static summaries, videos, and to an extent, long‑form articles), podcasts offer a conversational, human‑like delivery that may foster perceptions of coherence, trust, and social presence. Existing work in HCI, media and communication suggests that such qualities can meaningfully influence persuasion and attitude formation~\cite{lee2004presence, mcclung2010examining, matz2024potential,hackenburg2025levers}. Thus, podcasts provide a particularly rich test case for understanding how passive, narrative‑driven modalities interact with active search behaviors.
Recent preliminary work compared the impact of SERPs and AI-generated podcasts on users attitude change and found no significant differences~\cite{jwang_mscthesis}. However, as these two media converge in everyday information-seeking behavior, it becomes essential to explore how their interaction influences user attitudes, particularly in contexts involving controversial, value-laden, and debated topics~\cite{wang2024cognitively,rieger2024disentangling,rieger2024responsible}. This informs the core research gap and the overarching question that we aim to address in our work---\textit{how do the information mediums of present-day search engine result pages (SERPs) and AI-generated podcasts interact to shape opinions of users?} To this end, we aim to investigate the user attitudinal effects of consuming information via SERPs and AI-generated podcasts, focusing on how the sequence and modality of exposure shape user opinions. Thus, we aim to answer the following research questions:  

\begin{framed}
\begin{small}
\begin{itemize}[leftmargin=*]
    \item \textbf{RQ1}: How does information medium sequence (i.e., \textit{SERP-first} versus \textit{podcast-first}) influence user attitude change?
    \item \textbf{RQ2}: How does the effect of information medium sequence %(\textit{SERP-first} versus \textit{podcast-first}) 
    on user attitude change differ across viewpoint biases (supporting, opposing, neutral)? %of SERPs and the AI-generated podcasts?
    \item \textbf{RQ3:} How does topic controversiality influence user attitude change across segments within information medium sequences?
    %\item \textbf{RQ3}: How does the effect of information medium sequence %(\textit{SERP-first} versus \textit{podcast-first}) 
    %on user attitude change differ across topics with respect to their degree of controversiality?
    \item \textbf{RQ4}: How do individual differences in AI literacy, intellectual humility, need for cognition, and user engagement influence attitude change across segments within  medium sequences?
    %\item \textbf{RQ4}: How is the effect of information medium sequence %(\textit{SERP-first} versus \textit{podcast-first}) 
    %on users' attitude change influenced by their intellectual humility, AI literacy, need for cognition, and engagement?
\end{itemize}  
\end{small}
\end{framed}

We carried out a $2\times3\times2$ between-subjects exploratory study ($N=483$) to examine whether users who engage with SERPs before or after listening to an AI-generated podcast experience different levels of attitude change. We investigated whether and how the sequence effects (\textit{SERP-first} versus \textit{podcast-first}) vary across three different viewpoint biases embedded in the content---either \texttt{supporting}, \texttt{opposing}, or \texttt{neutral}, and across topics with varying degrees of controversiality (\textit{moderate} versus \textit{high}). Acknowledging that individual differences can play a significant role in how information is consumed and processed, we explored whether users'  AI literacy, intellectual humility, need for cognition, and their perceived user engagement can moderate the impact of information medium sequence on their attitude change. By incorporating these variables, our study aims to provide a nuanced understanding of how users engage with and are influenced by multimodal information ecosystems that are increasingly commonplace today.

\textbf{Original Contributions.}
 We found that users who consumed information in podcasts before SERPs exhibited a significantly higher attitude change. Our results suggest a significant interaction between time and viewpoint bias, such that users who encountered opposing viewpoints were less prone to attitude change, particularly when such information is reinforced in a consequent medium. We also found that moderately controversial topics corresponded to a significantly higher user attitude change compared to highly controversial topics. 
These key findings and others reported in the paper have important implications for the design of multimodal search systems and responsible content delivery systems. All relevant data, code, and supplementary material can be accessed publicly in an OSF repository for reproducibility.\footnote{\url{https://osf.io/bn2pd/overview?view_only=2146ca25613c40308d195a5924444326}} 

%% file: 2_RelatedWork.tex
\section{Background and Related Literature}

We position our work in the context of existing literature on (i) web search and opinion formation, (ii) search in the age of generative AI, and (iii) multimodal information access.

\subsection{Web Search and Opinion Formation}
\label{sec-search}
Web search engines are among the most widely used gateways to information, and the results they deliver have significant power to shape people's opinions and decisions~\cite{canca2022did,white2015belief,chacoma2015opinion,rieger2024responsible}.
When people search for information on debated topics, subjects of ongoing discussion that lack a clear consensus, different factors can hinder responsible opinion formation, which would require actively and thoroughly engaging with diverse viewpoints~\cite{rieger2024responsible}.   
For instance, users' attitudes can be shifted by viewpoint biased search results, when some viewpoints are overrepresented compared to others~\cite{draws2021not} or elements on the search engine result page like featured snippets~\cite{bink2023investigating}.
Further, user factors like strong pre-existing attitudes on debated topics were observed to be linked to decreased engagement with diverse viewpoints and a low likelihood of attitude change~\cite{rieger2024disentangling, wang2024cognitively}.

These observations highlight the limitations of traditional search interfaces in supporting responsible opinion formation. This raises the question of what alternative information access systems and interfaces that support users in engaging more thoroughly and critically with different viewpoints on debated topics could look like~\cite{rieger2024responsible, shah_envisioning_2024} and whether generative AI could serve as a tool that contributes to that goal.

\subsection{Search in the Age of Generative AI}

% immediate effects: Research into how people search (differently) with GenAI systems and how that affects search outcomes:
Generative AI, and Large Language Models in particular, have shifted many people's information behavior~\cite{zhou_understanding_2024, white2025information}.
Although the precise ways in which they influence information behavior and outcomes is still mostly unclear, recent research has offered some initial insights into the effects of LLM based chat interfaces. For instance, \citet{yang2025search+} compared information seeking for complex learning tasks with traditional search interfaces to search with chat interface and found that participants engaged less with traditional search results when they could use the chat interface, and that the chat interface improved immediate learning, yet not longer term retention. Building on these findings with a follow-up qualitative study ~\citet{mayerhofer_blending_2025} found that while the level of correct responses for complex information tasks found with the search and chat interface was comparable to that achieved with traditional search interfaces, participants often misplaced trust for ease-of-use, leading to high post-search confidence even if their answer was incorrect, particularly when using the chat feature.

%Risks of Gen AI in information access: 
While we are beginning to understand some of the immediate effects of widely available LLMs on information behavior and search outcomes, we still know little about longer term consequences of these technologies for individuals and society. Focusing on information access, such systems could, for instance, disrupt the information ecosystem, further increase marginalization and power concentration, and hamper innovation~\cite{mitra_sociotechnical_2025}.
% something on the searcher?

% potential of GenAI in information access
Thus, GenAI, and LLMs in particular, pose  risks to our information ecosystems, yet, they also have the potential to facilitate information access and support information seeking in various ways~\cite{white_advancing_2024, shah_envisioning_2024}. 
For instance, GenAI can enable multimodal ways to access and interact with information, thereby potentially improving engagement, learning outcomes, and overall accessibility~\cite{trippas_adapting_2025,shah_envisioning_2024, aliannejadi_interactions_2025,deldjoo_multimodal_2021}.

\subsection{Multimodal Information Access}
\label{sec-multimodal}
%(what is multimodality and what can it be good for) 
Multimodal interfaces enable different modes of interacting with a system, via multiple modes of input and output, such as text, speech, gesture, image, or videos, or with different interaction paradigms, such as query-response or multi-turn dialog~\cite{white2025panmodal}. 
Traditional search engine interfaces have offered information interactions mainly via a single modality: visually, with query-response interactions~\cite{white2025panmodal, deldjoo_multimodal_2021}. 
However, researchers who investigated alternatives modes of information presentation, and interfaces that support access through other modalities, have observed a range of benefits.
%
% multimedia
Beyond the web search context, researchers found that presenting people with multimedia content during learning tasks (e.g., audio, video, text) reduces cognitive load, improves information comprehension, and increases motivation~\cite{cong2022measurement, lauc2020effects, septiani2021development}.
\citet{moraes_contrasting_2018} observed that presenting participants with an instructional video combined with a subsequent phase of conducting a web search yielded better learning outcomes than search or instructional video alone. 

% conversational
The advances of generative AI facilitate multimodal approaches to presenting and accessing information that can support deeper engagement with complex information, for instance through conversational interactions~\cite{white2025panmodal}.
Compared to traditional, primarily visual web search interfaces conversational modes of information access enable more natural, accessible, and in the case of spoken conversational modes, screen-free ways of interacting with information~\cite{deldjoo_multimodal_2021}. They also allow for information to be presented not only as facts, but with a richer context, for instance with added narration, which might play an important role in improving human comprehension~\cite{sadirijavadi_unveiling_2024}.

Some of the advantages of conversational interactions could also apply to delivering information through AI generated podcasts, which can transform written text into audio content, potentially increasing engagement and accessibility~\cite{huffman2024enhancing, yahagi2024paperwave}.
In an educational setting, \citet{do2024paige} oberserved that students were more engaged by AI-generated podcasts compared to textbook reading, and that podcasts tailored to student profiles improved knowledge retention in some school subjects.

While AI generated podcasts may improve some aspects of information interaction, they also pose various risk, such as misleading content, misrepresentation of complex ideas, reduced critical engagement due to the passive nature of the interaction, or tone-introduced bias in user reliance. 
These concerns become particularly critical when the content presented can shape people's opinions and decisions, which can be the case with information on debated topics~\cite{rieger2024responsible}.
However, combining AI generated podcasts with web search might leverage some benefits of multimodal information interactions, such as promoting deeper understanding and greater user engagement.
To advance the understanding of the potential benefits and risks of AI generated podcasts, we explore how multimodal interactions with information on debated topics, combining AI-generated podcasts with traditional web search interfaces affect attitude change.

%% file: 3_Method.tex
\section{Method}

% - For attitude change research, the Limitations-Owning Intellectual Humility Scale (LOIHS) is generally preferable to broader intellectual humility scales. 

% Advantages: The LOIHS was specifically developed by Haggard et al. (2018) to find "middle ground between intellectual arrogance and intellectual servility", making it more nuanced for measuring attitude change processes.

% Attitude Change Relevance: Recent research shows IH is associated with "people's latitude of acceptance within a social judgment framework of attitude change", and the limitations-owning conceptualization directly captures the cognitive flexibility needed for attitude revision.

% Specificity for Change Processes: The limitations-owning approach defines intellectual humility as "owning one's knowledge limitations", which is precisely what's required for genuine attitude change - acknowledging that current beliefs might be incomplete or incorrect.

%\subsection{Experimental Setup}

\begin{table*}[htbp]
    \centering
    \captionsetup{justification=centering}
    \caption{The set of \textit{moderately controversial} (MC) and \textit{highly controversial} (HC) topics considered in our study.}
    \label{tab:topics}
    \rowcolors{2}{gray!10}{white}
    \scalebox{.75}{
    \begin{tabular}{@{}p{50mm}p{120mm}@{}}
        \toprule
        \textbf{Degree of Controversiality} & \textbf{Topic} \\
        \midrule
        \rowcolor{gray!20}
        \textbf{Moderately Controversial (MC)} 
        & Cell phone radiation is safe. (Cell Phone) \\
        & Social networking sites are good for our society. (Social Networks) \\
        & Obesity is a disease. (Obesity) \\        
        \midrule
        \rowcolor{gray!20}
        \textbf{Highly Controversial (HC)} & 
        The government should allow more refugees to resettle in the United States. (Immigration) \\
        & Permitless carry of guns should be legal. (Gun Control) \\
        & Abortion should be legal. (Abortion) \\
        \bottomrule
    \end{tabular}}
\end{table*}

To understand how different information mediums (SERPs and AI-generated podcasts) interact in sequence to shape the opinions of users, 
we carried out a $2\times3\times2$ {between-subjects} crowdsourcing study that was approved by our university ethics committee. We examined two sequences (\textit{SERP-first} versus \textit{Podcast-first}), across three levels of viewpoint bias (\texttt{supporting}, \texttt{neutral}, and \texttt{opposing}) and topics that varied on the degree of controversiality (\textit{moderately} controversial versus \textit{highly} controversial topics, as shown in Table \ref{tab:topics}). These topics were chosen based on prior literature that has explored the role of viewpoint biases in SERPs, user opinion formation, and attitude change~\cite{draws2021not,xu2021user,rieger2024disentangling}. For instance, topics were selected from ProCon~\cite{procon2023}, a resource that presents controversial
topics and related arguments, with varying levels of controversy---\textit{‘Should abortion be legal?’} (\textit{highly} controversial, \textit{i.e.,} people tend to have
strong attitudes) and \textit{‘Is obesity a disease?’} (\textit{moderately} controversial, \textit{i.e.,} people tend to have moderate attitudes). This allowed us to assess and compare the attitude change effects under different conditions.

\subsection{Materials}

We first describe the process of generating the content of SERPs and the AI-generated podcasts presented to participants.

\subsubsection{\textbf{SERPs}}
We constructed SERPs corresponding to each of the six topics as follows: 
\begin{itemize}[leftmargin=*, nosep]
    \item[(i)] For moderately controversial topics, we directly used the publicly available viewpoint annotated dataset released by %from the "search results annotated" dataset
\citet{draws2021not}\footnote{https://osf.io/v38c5}, which contains the URL, title, snippet, and annotated viewpoint (7-point scale, from "\textit{strongly opposing}" to "\textit{strongly supporting}") of the search results;
    \item[(ii)] For highly controversial topics, we identified corresponding relevant sources on ProCon,\footnote{https://www.britannica.com/procon} and employed NotebookLM\footnote{https://notebooklm.google/} to annotate the viewpoints represented by the sources in a cost-effective manner~\cite{wang2024human}. Following best practices~\cite{tornberg2024best} and to confirm the quality of the viewpoint annotations by NotebookLM---given that large language models can be prone to erroneous annotations~\cite{tan2024large}---we carried out a validation on a random subset with authors of this paper acting as experts. 
    An example prompt (\textbf{P1}) used for the viewpoint annotation process is presented below.
\end{itemize}

\scalebox{.8}{
\begin{promptbox}
\textbf{P1.} {Please rate the viewpoint of each source about the statement that "\textbf{Obesity is a disease.}" You have the following seven choices: \textbf{strongly opposing}, \textbf{opposing}, \textbf{somewhat opposing}, \textbf{neutral}, \textbf{somewhat supporting}, \textbf{supporting}, \textbf{strongly supporting}.} 
\end{promptbox}}

Apart from the retrieval and viewpoint annotation of the search results used in our study, we followed the ranking methodology employed by \citet{draws2021not} to create SERPs that are either \texttt{supporting}, \texttt{opposing}, or \texttt{neutral} with respect to the topic. 
We randomly sampled three “opposing”, two “somewhat opposing”, two “somewhat supporting”, and three “supporting” items from the search result items that were deemed relevant to a given topic. 
For the supporting condition, we used the ranking with extreme bias towards the supporting viewpoint. For the opposing condition, we used the ranking with extreme bias towards the opposing viewpoint. For the neutral condition, we used the ranking with little bias (wherein the supporting viewpoints and opposing viewpoints are interleaved, with half of the results containing bias for the opposing viewpoint and the other half for the supporting viewpoint). Consistent with existing practices around using AI overviews of SERPs (e.g., on Google search), and to help participants grasp key information from the SERPs, we generated and integrated a concise AI overview using the following prompt (\textbf{P2}):

\scalebox{.8}{
\begin{promptbox}
    \textbf{P2.} \textit{Imagine that a user inputs the query "\textbf{Is obesity a disease?}" into the search box of a search engine and receives these 10 sources in return. Now, please generate a short overview of the search results (the number of words should be less than 150).}
\end{promptbox}} 

%% Uncomment for pre-print version!
%Figure \ref{fig:serp} presents the SERP interface in our study, corresponding to the topic of ``\textit{Cell phone radiation is safe}.''

%% Uncomment for pre-print version!
% \begin{figure}[htbp] 
% \centering
% \includegraphics[width=.75\columnwidth]{Figures/SERP_cropped.jpg}
% \captionsetup{justification=centering}
% \caption{Screenshot of the SERP interface in our study demonstrating the AI overview and the search results.} 
% \label{fig:serp} 
% \end{figure}

\subsubsection{\textbf{AI Podcast}}

%\ujwal{Bookmark for my ongoing pass!}

Using the "Audio Overview" feature of NotebookLM, we generated podcasts reflecting different viewpoint biases based on the SERPs. The AI model was instructed to synthesize content from topic sources while adhering to a predefined bias. E.g.,:

\scalebox{.8}{
\begin{promptbox}
    \textbf{P3.} \textit{Focus on discussing whether obesity is a disease. Focus on the sources which hold the view that obesity is a disease.}\\
    \textbf{P4.} \textit{Focus on discussing whether obesity is a disease. Ensure a balance between the sources which hold the view that obesity is a disease and the sources which hold the view that obesity is not a disease.} \\
    \textbf{P5.} \textit{Focus on discussing whether obesity is a disease. Focus on the sources which hold the view that obesity is not a disease.} 
\end{promptbox}}

 %Example prompts are presented above.
Two authors of this paper independently assessed the alignment of the 18 AI-generated podcasts with the intended viewpoint biases (i.e., 6 topics with 3 different viewpoint biases each). Inter-rater reliability was assessed using Krippendorff’s $\alpha$ for ordinal data, which accounts for the magnitude of disagreements on an ordered scale. We found $\alpha$ = 0.896 (95\% CI: $0.78-0.97$), indicating very high agreement between raters, supporting the reliability of the embedded viewpoint biases in the AI-generated podcasts. 

%% Uncomment for pre-print version!
% \begin{figure}[htbp] 
% \centering
% \includegraphics[width=.75\columnwidth]{Figures/podcast_cropped.jpg}
% %\vspace{15pt}
% \captionsetup{justification=centering}
% \caption{Screenshot of the podcast interface in our study.}
% \label{fig:podcast}
% \end{figure}

\subsection{Measures}

\textbf{Attitude change.} We measure \textit{attitude change} as the main dependent variable in our study, drawing inspiration from prior works~\cite{draws2021not,rieger2024disentangling} that have captured participant attitudes before and after exposure to information (e.g., via SERPs). We captured the attitudes of participants corresponding to the different topics at three different points in our study---their pre-existing attitude on the topic ($pre$), their attitude after exposure to the first medium ($mid$), and after exposure to the second medium ($post$). For example, we used the following question: \textit{To what extent do you agree with the following statement? `Cell phone radiation is safe.'}, and capture responses on a 7-point Likert-scale \textit{-3: Strongly Opposing}---to---\textit{+3: Strongly Supporting}. This allows us to measure the participants' attitude change across two piecewise segments: (i) $\Delta_1$ from $pre$ to $mid$, and (ii) $\Delta_2$ from $mid$ to $post$. 

\textbf{Attitude strength.} 
\citet{luttrell2020attitude} have shed light on the difference between predictors and defining features of \textit{attitude strength}, which is defined as the durability and impact of an attitude. They synthesized existing literature, arguing that strength-related attitude attributes and attitude strength itself are independent. Therefore, we separately capture three key attributes that have been shown to be effective predictors of attitude strength: %as follows. %using the following questions. 
\begin{itemize}[leftmargin=*,nosep]
    \item[(i)] \textit{Importance} (i.e., how much a person perceives him or herself to care about a particular attitude)~\cite{eaton2008attitude,haddock1999forming,leeper2020more,holbrook2005attitude}; we captured importance at the pre-task stage using the following question: `\textit{How important is your attitude on this topic to you?},' on a 7-point Likert-scale (\textit{1: Not important at all}---to---\textit{7: Extremely important}).
    \item[(ii)] \textit{Elaboration} (i.e., the extent to which attitudes are formed through careful thinking about relevant information)~\cite{barden2014elaboration,petty2014elaboration,haugtvedt1994message,horcajo2016effects}; we captured elaboration using the following question: \textit{To what extent do you agree with the following statement? ``Your attitude on this topic is a result of careful thinking about relevant information.''} on a 7-point Likert-scale (\textit{1: Strongly Disagree}---to---\textit{7: Strongly Agree}) at three points in the study (at the pre-task stage, after exposure to the first medium, after exposure to the second medium). 
    \item[(iii)] \textit{Moralization} (i.e., the degree to which people perceive an attitude as being connected to core moral values) ~\cite{skitka2014psychological,skitka2014social,aramovich2012opposing,luttrell2016making}; we captured this at the pre-task stage using the following question: \textit{To what extent do you agree with the following statement? ``Your attitude on this topic is connected to your core moral values.''} on a 7-point Likert-scale (\textit{1: Strongly Disagree}---to---\textit{7: Strongly Agree}). 
\end{itemize}

\textbf{Belief strength.} We captured the \textit{belief strength} of participants in our study. This refers to how strongly a person holds a specific belief to be true or the degree of certainty that one has in the truth of a specific proposition ~\cite{egan1986concept,dennin2022relationship,vidigal2025measuring}.  We used the following two questions to capture belief strengths at three points in the study (at the pre-task stage, after exposure to the first medium, after exposure to the second medium) ; (i) \textit{How strongly do you believe your answer is based on knowledge rather than intuition?} on a 7-point Likert-scale (\textit{1: Strongly Disagree}---to---\textit{7: Strongly Agree}), and (ii) \textit{How certain are you about your beliefs on this topic?} on a 7-point Likert-scale from (\textit{1: Completely Uncertain}---to---\textit{7: Completely Certain}).

\textbf{AI literacy (AIL).} Given the context of our study and the real-world state of SERPs with embedded AI overviews and increasingly accessible AI-generated podcasts, we captured \textit{AI literacy} of the participants in our study as a potential moderator. To this end, we employed the 10-item short version of the Meta AI Literacy Scale (MAILS)~\cite{koch2024meta,carolus2023mails}. MAILS is a scale that helps us measure AI literacy and other psychological competencies %(problem-solving, learning, self-efficacy, persuasion literacy)
that are assumed to support successful interaction with AI. 

\textbf{Intellectual humility (IH).} Recognizing gaps in one’s knowledge and that one’s current beliefs might be incorrect can play an important role in shaping attitudes on different topics~\cite{rieger2024potential}.
\textit{Intellectual humility} is the ability
to recognize shortcomings or potential limitations in
one’s own point of view~\cite{whitcomb2017intellectual,porter2022predictors}. We employed the 12-item Limitations-Owning Intellectual Humility Scale (L-OIHS) since it allows for a relatively more nuanced interpretation and measurement of intellectual humility~\cite{haggard2018finding}.

\textbf{Need for cognition (NFC).} NFC is a psychological trait %that represents 
representing the extent to which individuals enjoy and engage in effortful cognitive activities~\cite{cacioppo1982need}. \citet{haugtvedt1992personality} found that NFC moderates the persistence and resistance of attitude and belief changes among individuals. ~\citet{wu2014using} found that users with a higher NFC demonstrated different search behavior and different interactions with SERPs than those with a lower NFC. Given the wealth of evidence in existing literature that has demonstrated how NFC influences attitude change and user interactions with information, we measured users' NFC using the \textit{NFC-6} scale  developed by \citet{lins2020very}. %\ujwal{TODO: Carry out an exploratory analysis comparing high-NFC versus low-NFC users. OR NOT.}

\textbf{User engagement (UE).} This relates to the extent to which a user is actively involved with a digital experience, that has been shown to be important in the context of web search~\cite{zhuang2017understanding}. %. \ujwal{Add the standard IR-related motivation for capturing user engagement here.}
We measured user engagement using the UES-SF scale developed by~\citet{o2018practical} and widely used in the context of user interactions with digital media.

\subsection{Participants}

%\textbf{Simulation-based power analysis.} To estimate the required sample size for our study, we carried out a simulation-based power analysis considering a piecewise difference-in-differences mixed model (described below). This model includes repeated measures at three time points ($pre$, $mid$, $post$), fixed effects for time and four covariates (viz., AI literacy, intellectual humility, need for cognition, and user engagement), and random intercepts and slopes for participants, as well as random intercepts for topics. We simulated data with a small-to-medium effect size (Cohen’s $f = 0.10-0.25$), while considering noise in the outcome variable, and the influence of covariates and their interactions.
%The simulation was run across sample sizes ranging from 25 to 30 participants per topic and for each viewpoint bias, balanced across the sequence modality (total $N = 450-540$), using 500 iterations per condition. Our results showed that, under these conservative assumptions, the model maintained adequate power ($\geq 0.80$) to detect changes in attitude over time. 

%\noindent\textbf{Crowdsourcing.} %Based on a simulated power analysis and accounting for potential participant exclusion, 
Based on estimates from a simulated power analysis and accounting for potential participant exclusion, we recruited 550 participants through the Prolific crowdsourcing platform.\footnote{https://www.prolific.com/} All participants were paid an hourly wage of 9 GBP per hour (per the guidelines for a `\textit{good}' payment on Prolific). We restricted participation to workers in the age group of 18 and 50, who had successfully completed over 40 tasks and maintained an approval rate of over 90\%. The gender ratio was balanced in terms of the female-male ratio (1:1), and due to the topics considered---which are commonly debated in the USA---we restricted participation to workers from USA. To ensure participant reliability, we included 5 questions (that doubled as attention checks in the study) that participants encountered after each medium, and we rewarded participants with bonuses of 0.25 GBP for each correct answer. This design also allowed us to incentivize genuine interaction with the SERPs and the podcast, increasing the external validity of data gathered in our study. 

\subsection{Procedure}

Participants were first informed about the scope of the study, the data that would be gathered, and provided consent to proceed. They were then asked to respond to questions in a pre-task questionnaire, including their existing attitude on a given topic ($pre$), the three predictors of attitude strength, belief strength, AI literacy, intellectual humility, and need for cognition. Next, they were asked to progress through a tutorial that allowed them to familiarize themselves with the task and the interface. After the tutorial, participants moved on to the main phase of the study, where they were asked to engage with information related to the given topic. Depending on the experimental condition, they were either exposed to the SERPs first or the podcast first. Participants were asked to engage with the search results (e.g., by clicking on them to retrieve the linked webpages) and listen to the podcasts, and proceed once they were satisfied with the SERPs they consumed %\revise{do we mean, search results they clicked on here?} 
(or the podcast they heard) on the given controversial topic. Following the first medium and before they are directed to the second, we capture the attitude of participants on the topic once again ($mid$), along with their attitude strength (elaboration), and belief strength. After the second medium, we capture the attitude of participants on the topic ($post$), alongside their attitude strength (elaboration), and belief strength. Participants were finally directed to a post-task questionnaire where we captured their perceived engagement and gathered open-ended responses comparing the two media.   %\footnote{Note that participants had to click on at least two SERPs before they could proceed, and they were also required to listen to at least} 

\subsection{Piecewise DiD Mixed Model}\label{sec:mixed_model}

%\subsubsection*{\textbf{Data Filtering}}
\label{sec:attention}

To collectively address our research questions, we adopt a piecewise difference-in-differences (DiD) mixed model. This can be formally represented as follows: 

\[
\begin{aligned}
Y_{it} &= \beta_0 
+ \beta_1\,s1_{t} + \beta_2\,s2_{t}
+ \boldsymbol{\beta}_3^\top \mathbf{X}_i
+ \boldsymbol{\beta}_4^\top \big(s1_t \cdot \mathbf{X}_i\big)
+ \boldsymbol{\beta}_5^\top \big(s2_t \cdot \mathbf{X}_i\big) \\
&\quad + u_{0i} + u_{1i}\,s1_t + u_{2i}\,s2_t
+ v_{0,\text{topic}(i)} + \varepsilon_{it},
\end{aligned}
\]

where, $Y_{it}$ represents the attitude of participant $i$ at time $t$ $\in$ $\{pre,mid,post\}$;
%\indent\indent\indent
$\beta_0$ represents the fixed intercept (i.e., the expected attitude score at $pre$);
%\indent\indent\indent
$s1_{t}=1[t\in{mid,post}]$ (piecewise time indicator which captures $\Delta_1$, the first segment change from $pre$ to $mid$);
%\indent\indent\indent
$s2_{t}=1[t={post}]$ (piecewise time indicator which captures $\Delta_2$, the second segment change from $mid$ to $post$); 
%\indent\indent\indent
$\beta_1$ represents the average change in attitude after the first medium, and $\beta_2$ represents the additional change after the second medium;
%\indent\indent\indent 
${X}_i$ represents the vector of between-subjects predictors: the sequence (\texttt{SERP-first} or \texttt{Podcast-first}), viewpoint bias (\texttt{supporting}, \texttt{opposing}, \texttt{neutral}), controversiality (\texttt{moderate}, \texttt{high}), and moderators (\textit{AI literacy}, \textit{need for cognition}, \textit{intellectual humility}, \textit{user engagement});
%\indent\indent\indent
$\beta_3^\top {X}_i$ represents the main effects of predictors on the baseline attitude $(pre)$;
%\indent\indent\indent
$\beta_4^\top \big(s1_t \cdot \mathbf{X}_i\big)$ represents the interactions of the first-segment change $\Delta_1$ with predictors;
%\indent\indent\indent
$\beta_5^\top \big(s2_t \cdot \mathbf{X}_i\big)$ represents the interactions of the second-segment change $\Delta_2$ with predictors;
%\indent\indent\indent
$u_{0i}$ represents the participant-specific intercept (baseline differences), $u_{1i}$ represents the participant-specific deviation in $\Delta_1$, and $u_{2i}$ represents the participant-specific deviation in $\Delta_2$ for participant $i$;
%\indent\indent\indent
$v_{0,\text{topic}(i)}$ represents the random intercept for topic to account for topic-level differences

%% file: 4_Results_new.tex
\section{Results and Analysis}

\textbf{Descriptive Insights.}
To ensure data quality, we assessed participants' attention levels using a heuristic composed of: (i) time spent on the SERPs/podcast, and (ii) accuracy on attention check questions related to the content. These were chosen since the time spent and accuracy on attention check questions together have been shown to serve as a reasonably good proxy for reliability in similar crowdsourced studies~\cite{gadiraju2018analyzing,rieger2024disentangling}. For each experimental condition, we first computed the \textit{median time spent} and the \textit{median accuracy rate} across all participants. We then computed an \textit{attention score} for each participant and each medium (SERP, podcast) as follows:
% %
% \[
% \text{Attention Score} = \frac{\text{Time Spent}}{\text{Median Time}} + \frac{\text{Accuracy Rate}}{\text{Median Accuracy Rate}}
% \]
% %
%
\[
\scalebox{0.95}{$
\text{Attention Score} = \frac{\text{Time Spent}}{\text{Median Time}} +
\frac{\text{Accuracy Rate}}{\text{Median Accuracy Rate}}
$}
\]

Participants whose attention scores were below 1 in both mediums were excluded from subsequent analyses.
%\revise{How many participants were excluded? Reveiwer 1 comment}
%This threshold ensures that only participants who demonstrated minimal engagement and comprehension across sessions were filtered out.
%
%We filtered out the data from participants who were deemed to be unreliable. This was estimated using  final number of 
%
The resulting distribution of participants in each group can be seen in Table \ref{tab:number}. %Overall intellectual humility (mean = 2.70, sd = 1.08), AI literacy (mean = 6.78, sd = 1.75), need for cognition (mean = 1.58, sd = 1.58), preexisting viewpoint (mean = 0.08, sd = 11.78), and user engagement (mean = 2.42, sd = 0.55) were moderate. The standard deviation of preexisting viewpoint is relatively higher than others, it's because we calculate the attitude by multiplying participants' agreement by their strength belief.

\begin{table}[h]
    \centering
    \renewcommand{\arraystretch}{1.1}
    \rowcolors{2}{gray!10}{white}
        \caption{Number of reliable (and excluded) participants  in each group considered in our analysis, split by topic (degree of controversiality) and across the three viewpoint biases.}
    \label{tab:number}
    \scalebox{.75}{
    \begin{tabular}{@{}lrrr@{}}
        \hline
        \textbf{Topic} & \textbf{Supporting} & \textbf{Neutral} & \textbf{Opposing} \\
        \midrule
        Cell Phone (MC)       & 27 (7) & 28 (6) & 27 (7) \\
        Social Networks (MC)      & 26 (4) & 30 (0) & 29 (1) \\
        Obesity (MC)     & 26 (4) & 29 (1) & 26 (4)\\
        Immigration (HC) & 26 (4) & 28 (2) & 29 (1)\\
        Gun Control (HC)    & 26 (4) & 24 (6) & 27 (3)\\
        Abortion (HC)  & 25 (5) & 25 (5) & 25 (5)\\
        \hline
        \rowcolor{gray!20}
        \textbf{Total} & \textbf{156} (28) & \textbf{164} (20) & \textbf{163} (21)\\
        \hline
    \end{tabular}}
%\captionsetup{justification=centering}
\end{table}

Note that data %further exploratory analysis 
pertaining to the attitude strength predictors (importance, elaboration, moralization) and belief strength are included in the online repository, beyond the scope of our primary analysis presented in the paper.

We analyzed attitude change among users across the different conditions to identify and categorize the instances where user attitudes \textit{flipped} (i.e., they changed to the opposing view), \textit{strengthened} (i.e., user attitudes were further reinforced), \textit{weakened}, or remained \textit{unchanged}. Table \ref{tab:attitude_change_medium} presents the distribution of attitude change outcomes based on the sequence medium. We found that the majority of users demonstrate unchanged attitudes, followed by nearly one-fourth of the users demonstrating a strengthening of attitudes, and the remaining users either reported flipped attitudes or weakened attitudes. This general trend was consistent across both sequences of \textit{Podcast-first} and \textit{SERP-first}. 

\begin{table}[h]
\centering
%\captionsetup{justification=centering}
\caption{Distribution of attitude change outcomes among users by sequence medium (\textit{Podcast-first} versus \textit{SERP-first}).}
\label{tab:attitude_change_medium}
\setlength{\tabcolsep}{10pt}
\renewcommand{\arraystretch}{1.15}
\rowcolors{2}{gray!10}{white}
\scalebox{.75}{
\begin{tabular}{l l r}
\toprule
\textbf{Sequence Medium} & \textbf{Outcome} & \textbf{\#Users (\%)} \\
\midrule
\textbf{Podcast-first} (\textit{243}) & Flipped      & \(43\) (17.7\%) \\
                                       & Strengthened & \(59\) (24.3\%) \\
                                       & Unchanged    & \(101\) (41.6\%) \\
                                       & Weakened     & \(40\) (16.5\%) \\
\midrule
\textbf{SERP-first} (\textit{240})    & Flipped      & \(37\) (15.4\%) \\
                                       & Strengthened & \(62\) (25.8\%) \\
                                       & Unchanged    & \(99\) (41.2\%) \\
                                       & Weakened     & \(42\) (17.5\%) \\
\bottomrule
\end{tabular}}
\end{table}

We also explored the distribution of attitude change outcomes among users across topics corresponding to a \textit{high} versus \textit{moderate} degree of controversiality (Table \ref{tab:attitude_change_controversiality}). Once again, we found similar trends in the outcomes. %Further distributions of attitude change outcomes among users across different topics and viewpoint biases are included in the online repository.    

\begin{table}[htbp]
\centering
\caption{Distribution of attitude change outcomes among users by degree of controversiality (\textit{High} versus \textit{Moderate}).}
\label{tab:attitude_change_controversiality}
\setlength{\tabcolsep}{10pt}
\renewcommand{\arraystretch}{1.15}
\rowcolors{2}{gray!10}{white}
\scalebox{.75}{
\begin{tabular}{l l r}
\toprule
\textbf{Controversiality} & \textbf{Outcome} & \textbf{\#Users (\%)} \\
\midrule
\textbf{High} (\textit{235})      & Flipped      & \(34\) (14.5\%) \\
                                  & Strengthened & \(59\) (25.1\%) \\
                                  & Unchanged    & \(107\) (45.5\%) \\
                                  & Weakened     & \(35\) (14.9\%) \\
\midrule
\textbf{Moderate} (\textit{248})  & Flipped      & \(46\) (18.5\%) \\
                                  & Strengthened & \(62\) (25.0\%) \\
                                  & Unchanged    & \(93\) (37.5\%) \\
                                  & Weakened     & \(47\) (19.0\%) \\
\bottomrule
\end{tabular}}
\end{table}

\textbf{Model Overview.} Our model incorporated random intercepts for both participants and topics to account for unobserved heterogeneity in baseline attitudes of users. The participant-level variance was found to be  $\sigma^2$ \textit{= 1.55}, indicating substantial individual differences in initial attitudes. This corroborates the importance of accounting for such variability while modeling attitudes longitudinally. This random effect captures how participants differ in their baseline responses, independent of the fixed predictors. Similarly, the topic-level variance was found to be $\sigma^2$\textit{= 1.418}, reflecting meaningful variation in users' attitudes across different topics. This supports the inclusion of topic as a random effect, acknowledging that some topics inherently elicit stronger or weaker attitudes regardless of the medium sequence or the viewpoint bias.

Overall, these random effects enhance our model’s robustness by allowing for more accurate estimation of fixed effects while controlling for nested data structures and contextual variability.

%\ujwal{TODO: Full pass to clarify results and explicitly align with the model in 3.5.}

\subsection{RQ1: Effect of information medium sequence on attitude change}

To examine how various predictors influenced users’ attitude changes ($\Delta_1$ and $\Delta_2$), we fit a piecewise difference-in-differences mixed-effects model. The fixed effects component of the model revealed that the sequence of information exposure significantly impacted user attitudes. Per our model described in Section~\ref{sec:mixed_model}, this corresponds to $\boldsymbol{\beta}_3$, which captures the main effect of medium sequence on baseline attitude ($pre$), and through interaction terms in $\boldsymbol{\beta}_4$ and $\boldsymbol{\beta}_5$ for segment-specific changes ($\Delta_1$ and $\Delta_2$). For RQ1, we focus on the main effect of sequence on users' attitude change.

Users who encountered SERPs first exhibited a lower  attitude change compared to those who engaged with podcasts first ($\beta$\textit{ = --0.376, p = .034; 95\% CI [-0.72, -0.03]}). This negative coefficient suggests that, holding other factors constant, beginning with SERPs first led to lower attitude change scores relative to starting with podcasts. 
The standardized coefficient for medium sequence was $\beta^\ast=-0.189$, indicating a small-to-moderate effect size. The unique contribution of medium sequence to variance explained was partial $R^{2}=.0031$, suggesting that while statistically significant, the effect accounts for about 0.3\% of residual variance. Cohen's $d$ for the contrast between SERP-first and Podcast-first was $d=-0.10$, reinforcing that the practical impact is small but meaningful in the context of attitude change. Interactions with segments were not significant, indicating that medium sequence primarily influenced baseline attitudes rather than segment-specific changes.

\textbf{Key takeaway (1)}: Information modality (\textit{i.e.,} SERPs or podcasts) and order of information presentation can shape initial impressions and subsequent attitude formation when these mediums are consumed in sequence.

\subsection{RQ2: Does the effect of sequence on attitude change differ across viewpoint biases?}

On exploring the effect of medium sequence on attitude change across different viewpoint biases, we found that the opposing viewpoint bias had a negative but non-significant main effect on users attitude change ($\beta=-0.304, p =.167,  \beta^\ast=-0.153$), while supporting viewpoint bias was negligible ($\beta=-0.076, p=.731, \beta^\ast=-0.038$.)
%On exploring the effect of medium sequence on attitude change across different viewpoint biases, we found that 

The interaction between time and viewpoint bias was significant in the second segment ($mid$$\to$$post$) with a small effect %Users who encountered opposing viewpoints exhibited significantly lesser attitude change than those who encountered supporting viewpoints in the $mid$$\to$$post$ segment 
($\beta=-0.297, p=.039;$ \textit{95\% CI [-0.59, -0.02]}, standardized $\beta^\ast=-0.15$, partial $R^2=.003$). This interaction explains 0.3\% of the residual variance, which is statistically significant but practically small. Because higher attitude change scores indicate a favorable change in attitude toward a given topic, a negative coefficient means that attitudes became less favorable (i.e., more negative) during this segment. Thus, when content was opposing, the additional change from $mid$$\to$$post$ was more negative compared to neutral bias. We did not find any significant interactions between medium sequence and viewpoint bias in the first segment. %($pre$$\to$$mid$). 

\textbf{Key takeaway (2):} During the second segment ($mid$$\to$$post$), users who encountered viewpoint opposing information became less favorable toward the topic compared to those exposed to neutral information, indicating that viewpoint opposing content drives a small but meaningful negative shift in user attitudes.

\begin{table}[ht]
\centering
\rowcolors{2}{gray!10}{white}
\caption{Differences in users' attitude change across viewpoint biases and sequences, relative to the neutral baseline.}
\label{tab:attitude_change_viewpoint_biases}
\scalebox{.75}{
\begin{tabular}{llccc}
\toprule
\textbf{Viewpoint Bias} & \textbf{Sequence} & \textbf{Difference} & \textbf{95\% CI} & \textbf{p-value} \\
\midrule
\multicolumn{5}{c}{\textbf{Attitude change from $pre$$\to$$mid$ ($\Delta_1$ = mid $-$ pre)}} \\
Supporting & SERP-first & $-0.163$ & [$-0.582$, $0.291$] & 0.46 \\
Opposing & SERP-first & $+0.350$ & [$-0.224$, $0.851$] & 0.20 \\
Supporting & Podcast-first & $+0.641$ & [$0.024$, $1.245$] & \textbf{0.04} \\
Opposing & Podcast-first & $-0.440$ & [$-1.003$, $0.118$] &  0.13\\
\midrule
\multicolumn{5}{c}{\textbf{Attitude change from $mid$$\to$$post$ ($\Delta_2$ = post $-$ mid)}} \\
Supporting & SERP-first & $+0.174$ & [$-0.233$, $0.57$] & 0.40 \\
Opposing & SERP-first & $-0.510$ & [$-0.931$, $-0.117$] & \textbf{0.02} \\
Supporting & Podcast-first & $-0.282$ & [$-0.631$, $0.01$] & 0.09 \\
Opposing & Podcast-first & $-0.166$ & [$-0.579$, $0.201$] & 0.39 \\
\bottomrule
\end{tabular}}
\end{table}

To examine whether viewpoint bias alters attitude change within each medium sequence, we compared segment changes for content with supporting and opposing biases against neutral content (the baseline) within sequence (\textit{SERP‑first} vs. \textit{Podcast‑first}) using bootstrap contrasts. Table \ref{tab:attitude_change_viewpoint_biases} presents the sequence differences in segment-specific attitude change ($\Delta_1$ and $\Delta_2$) across viewpoint biases. %, using the neutral viewpoint as the baseline.
A positive difference here indicates that the viewpoint bias condition produced a greater increase (or less decrease) in users' attitude (\textit{i.e.,} indicating a more favorable shift in attitude toward the topic than the neutral condition for that segment). A negative difference indicates a greater decrease (or a smaller increase) in users' attitude than neutral (\textit{i.e.,} indicating a relatively less favorable shift in attitude toward the topic). %Because higher attitude scores indicate more favorable attitudes toward the topic, negative differences imply becoming less favorable (i.e., more negative) relative to neutral, and positive differences imply becoming more favorable relative to neutral.
We found a statistically significant effect for the \textit{Podcast-first} sequence in $\Delta_1$ in the supporting viewpoint bias condition; (95\% CI: [0.024, 1.245], $p=.04$), indicating that supporting content led to a larger increase in user attitudes favorably towards a topic than neutral content during the first segment.
%the \textit{Podcast-first} sequence leads to a greater attitude change than \textit{SERP-first} sequence when supporting viewpoints are encountered by users. 
In addition, we found a significant effect for the \textit{SERP-first} sequence in $\Delta_2$ in the opposing viewpoint bias case; (95\% CI: [-0.931, -0.117], $p = .021$), indicating that exposure to viewpoint opposing information led participants to become less favorable toward the topic relative to those exposed to neutral information.
%the \textit{SERP-first} sequence leads to a lower attitude change than the \textit{Podcast-first} sequence when opposing viewpoints are encountered by users. 

\begin{figure}[!h]
    \centering
    \includegraphics[width=\columnwidth]{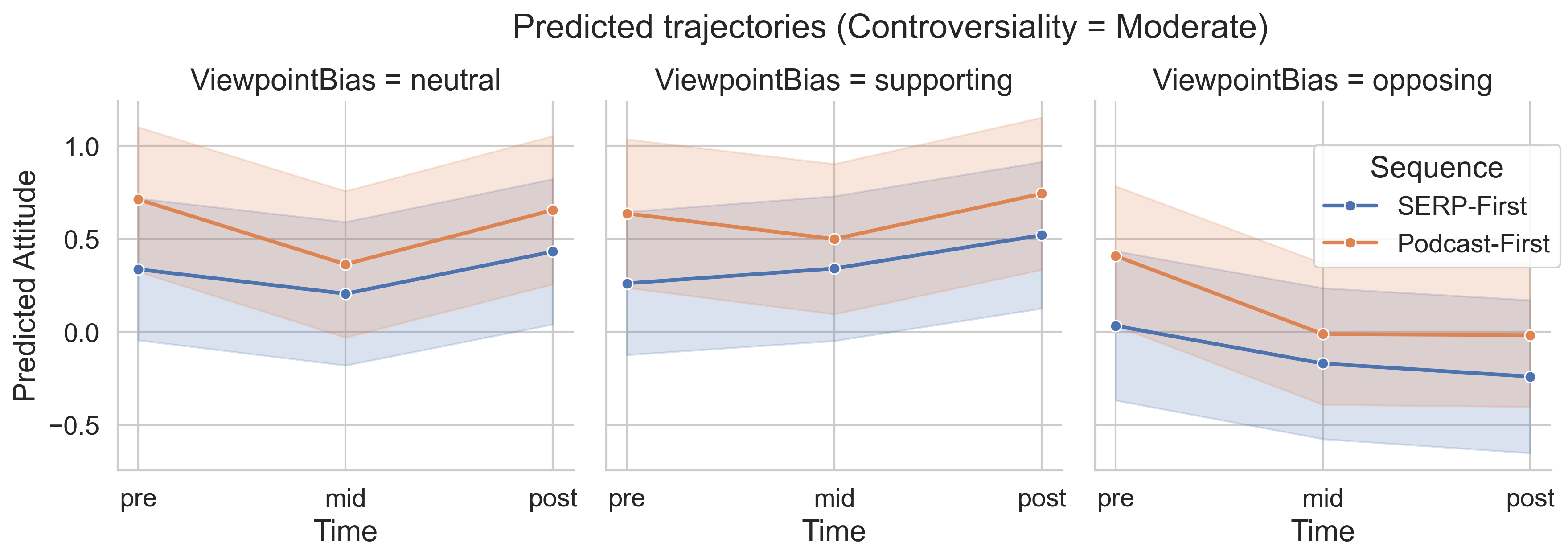}    
    % \captionsetup{justification=centering}
    \caption{Predicted attitude trajectories for each viewpoint bias across the three time points ($pre$, $mid$, $post$) for moderately controversial topics. Shaded regions represent 95\% CIs.}
    \label{fig:RQ1_1}
\end{figure}

\begin{figure}[!h]
    \centering
    \includegraphics[width=\columnwidth]{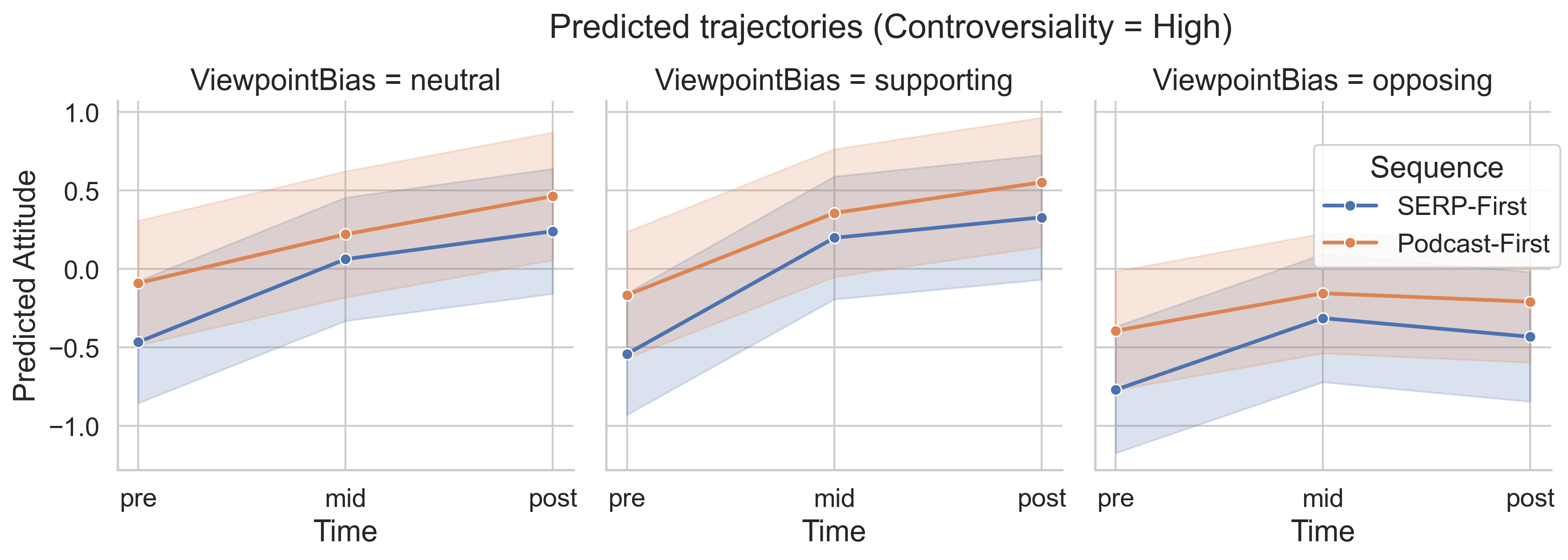}    
    % \captionsetup{justification=centering}
    \caption{Predicted attitude trajectories for each viewpoint bias across the three time points ($pre$, $mid$, $post$) for highly controversial topics. Shaded regions represent 95\% CIs.}
    \label{fig:RQ1_2}
\end{figure}

In the Figures~\ref{fig:RQ1_1} and \ref{fig:RQ1_2} we plot the predicted attitude trajectories using the proposed piecewise difference-in-differences mixed effects model across three different time points (viz. $pre$, $mid$, and $post$) where user attitudes on the topics were captured in our study. 

Figure~\ref{fig:RQ1_1} corresponds to moderately controversial topics and Figure~\ref{fig:RQ1_2} to highly controversial topics, respectively. Addressing RQ2, these plots contrast the sequence of mediums; \textit{SERP-first} versus \textit{Podcast-first}. These trajectory plots reveal distinct order effects across the different viewpoint biases.

\textbf{Key Takeaway (3)}: Within the SERP‑first sequence, opposing content produced a significantly more negative attitude change in the second segment ($\Delta_2$) relative to neutral, indicating that attitudes became less favorable from $mid$$\to$$post$ when the information was opposing. Within the Podcast‑first sequence, supporting content produced a significantly more positive attitude change in the first segment ($\Delta_1$) relative to neutral, indicating a larger gain in favorability from $pre$$\to$$mid$ when the information was supportive. %All other within-sequence differences were not statistically non-significant.

\subsection{RQ3: Does topic controversiality influence users attitude change across segments within information medium sequences?}%Does the effect of sequence on attitude change differ across topics with varying degrees of controversiality?}

\begin{figure}[h]
    \centering
    \includegraphics[width=.5\textwidth]{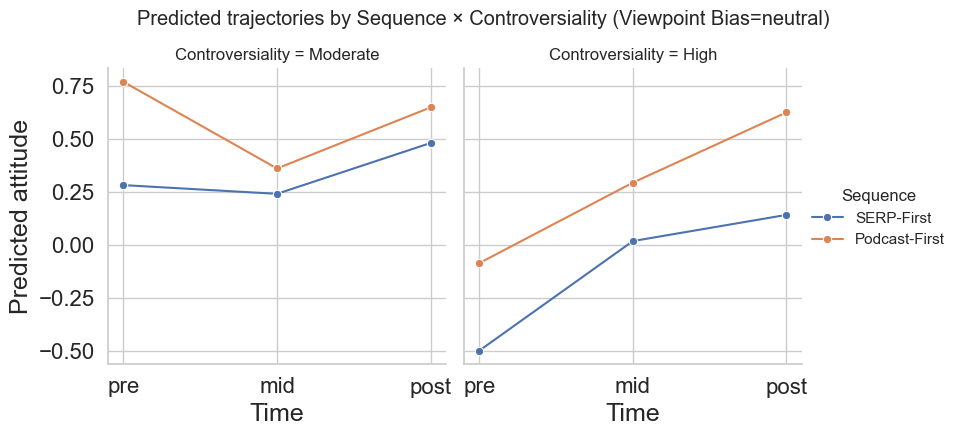}    
    % \captionsetup{justification=centering}
    \caption{Predicted attitude trajectories for topics with varying controversiality (\textit{moderate}, \textit{high}) across the time points ($pre$, $mid$, $post$), separated by the medium sequence (\textit{SERP-first}, \textit{Podcast-first}), and with viewpoint bias fixed to neutral.}
    \label{fig:RQ2_trajectories}
\end{figure}

Figure~\ref{fig:RQ2_trajectories} illustrates the predicted attitude trajectories for topics classified as \textit{Moderately} versus \textit{Highly} controversial, holding the neutral viewpoint bias constant. %For moderately and highly controversial topics, both sequences produced similar overall patterns, with small changes across segments.

%We tested whether the effect of information medium sequence (\textit{Podcast-first} versus \textit{SERP-first}) differed across topics by controversiality (\textit{high} versus \textit{moderate}) using our proposed mixed-effects model. %The model demonstrated a strong main effect of controversiality, with moderately controversial topics associated with higher attitude changes. 

We found that the main effect of controversiality was significant-- moderately controversial topics were associated with more favorable baseline attitudes compared to highly controversial topics ($\beta=0.803,p<.001$, \textit{95\% CI [0.457,1.149]}), with a standardized coefficient of $\beta^\ast=0.405$ and partial $R^2=.0143$. %This means that at the pre-exposure stage, participants were more favorable toward moderately controversial topics than highly controversial ones.
%We found that moderately controversial topics led to a significantly higher user attitude change in comparison to highly controversial topics ($\beta$ \textit{= 0.803, p<.001; 95\% CI [0.457, 1.149]}), indicating that users may react more strongly to topics perceived as relatively less controversial. 
The interaction between the degree of controversiality and $\Delta_1$ ($pre$$\to$$mid$) was significant with a moderate effect ($\beta=-0.660, p<.001$; $\beta^\ast=-0.333$, partial $R^2=0.127$). Since higher attitude scores indicate more favorable attitudes towards topics, this negative coefficient indicates that moderately controversial topics corresponded to a smaller increase (or a greater decrease) in favorability during the first segment ($pre$$\to$$mid$) compared to highly controversial topics. %This suggests that initial exposure to such topics may reinforce existing attitudes rather than shift them. No significant interaction was found in the second segment from $mid$$\to$$post$ ($\beta$ \textit{= 0.049, p=.673}).means 
This interaction explains about 1.3\% of residual variance, which is small but meaningful in the context attitude change.

\textbf{Key Takeaway (4):} Users reported more favorable attitudes toward moderately controversial topics to begin with, but these topics corresponded to less positive change (or more negative change) during the first segment of the medium sequence compared to highly controversial topics. In the second segment of the medium sequence, we found that controversiality did not influence attitude change significantly.

\subsection{RQ4: Do individual differences influence attitude change within medium sequences?}

As reported in the Table \ref{tab:sequence_moderators} below, we found that interactions involving all the individual difference moderators considered in our study (AI literacy, intellectual humility, need for cognition, and user engagement) were not statistically significant across segments.

\begin{table}[h]
\centering
\caption{Non-significant interactions between sequence and individual moderators over time.}
\label{tab:sequence_moderators}
\rowcolors{2}{gray!10}{white}
\scalebox{.8}{
\begin{tabular}{lcc}
\toprule
\textbf{Interaction Term} & \textbf{Coefficient ($\beta$)} & \textbf{\textit{p}-value} \\
\midrule
$s_1\times$ AI Literacy & 0.014 & .784 \\
$s_2\times$ AI Literacy & 0.004 & .915 \\
$s_1\times$ Intellectual Humility & 0.023 & .768 \\
$s_2\times$ Intellectual Humility & 0.051 & .389 \\
$s_1\times$ Need for Cognition & $-0.040$ & .496 \\
$s_2\times$ Need for Cognition & $-0.043$ & .327 \\
$s_1\times$ User Engagement & 0.231 & .138 \\
$s_2\times$ User Engagement & 0.006 & .961 \\
\bottomrule
\end{tabular}}
\end{table}

\begin{table*}[t]
\centering
\caption{Qualitative comparison of podcasts, SERPs, and their combination synthesized through participants' feedback.}
\label{tab-reasons}
\setlength{\tabcolsep}{4pt}
\scalebox{.825}{
\begin{tabular}{p{0.1\textwidth}| p{0.68\textwidth} | p{0.25\textwidth}}
\hline
\textbf{Medium} & \textbf{Advantages} & \textbf{Limitations} \\
\hline\hline

\multirow{4}{*}{\textbf{Podcast}} &
\begin{itemize}[leftmargin=*, nosep]
    \item \textbf{Engagement:} More engaging and easier to follow, especially for auditory learners.
    \item \textbf{Convenience:} Allows multitasking (e.g., listening while driving or cooking).
    \item \textbf{Narrative clarity:} Simplifies complex topics through conversational tones and storytelling.
    \item \textbf{Retention:} Improves memorization due to emotional and contextual delivery.
\end{itemize}
&
\begin{itemize}[leftmargin=*, nosep]
    \item Lack of source verification.
    \item Potential bias or repetitive content.
\end{itemize}
\\
\hline

\multirow{3}{*}{\textbf{SERP}} &
\begin{itemize}[leftmargin=*, nosep]
    \item \textbf{Depth and diversity:} Provides access to multiple sources, studies, and viewpoints.
    \item \textbf{Fact-checking:} Enables users to verify claims and explore primary sources.
    \item \textbf{Control:} Users can skim or dive deep into specific details at their own pace.
\end{itemize}
&
\begin{itemize}[leftmargin=*, nosep]
    \item Overwhelming volume of information.
    \item Time-consuming to sift through.
\end{itemize}
\\
\hline

\multirow{4}{*}{\textbf{Combination}} &
\begin{itemize}[leftmargin=*, nosep]
    \item \textbf{Comprehensive understanding:} Combines the depth of SERP with the accessibility of podcasts.
    \item \textbf{Balanced learning:} SERP offers factual rigor, while podcasts provide narrative context.
    \item \textbf{Critical thinking:} Encourages cross-referencing and reduces bias.
\end{itemize}
&
\begin{itemize}[leftmargin=*, nosep]
    \item Requires more time and effort to engage with both mediums.
\end{itemize}
\\
\hline
\end{tabular}
}
\end{table*}

%% file: 5_discussion.tex
\section{Discussion, Implications, and Future Work}

Major search platforms today are increasingly integrating AI‑generated audio summaries and podcast‑like outputs into search experiences. Tools such as NotebookLM, Perplexity's audio mode, and Google's experimental ``\textit{Listen}'' features illustrate a trend toward audio‑augmented search, where users can move fluidly between traditional search results and AI‑generated audio. Switching between SERPs and AI-generated podcasts reflects an emerging interaction pattern in modern search ecosystems. 

%Inspired by this, we investigated how AI-generated podcasts combined with interactions with SERPs shape user opinions, focusing on the effects sequence (\textit{SERP-first}, \textit{podcast-first}), across three different viewpoint biases embedded in the content (\texttt{supporting}, \texttt{opposing}, or \texttt{neutral}), and across two levels of controversiality (\textit{moderate}, \textit{high}).

Our work is situated in this emerging and important context. In our study addressing the interaction between mediums of SERPs and podcasts, we observed that participants who engaged with the podcasts first showed increased attitude change compared to those who engaged with the SERPs first (\textbf{RQ1}). 
This observed ability of AI-generated podcasts to shift attitudes more easily than web search mirrors the duality of benefits and risks of the podcast medium, promoting deeper engagement with different viewpoints but also posing a risk of undue influence on beliefs and the potential spread of misinformation~\cite{jacobson_podcast_2021, pathiyan2024everything}.
Our results demonstrate that exposure to opposing viewpoints can shift user attitudes less favorably towards a topic after initial exposure, potentially reflecting reactance, conflict, or belief consolidation processes in the second sequence medium (\textbf{RQ2}). %Participants who were exposed to attitude-opposing content were less likely to change their attitudes than those exposed to neutral attitude-supporting content (\textbf{RQ2}). 
%This suggests that these participants may have experienced a backlash effect whereby exposure attitude-opposing information causes them to become more entrenched in their opinions~\cite{nyhan2010corrections}.
Similar observations have been made in the context of web search on debated topics~\cite{rieger2024disentangling}.
%
%We found that attitude change varied between topics of different degrees of controversiality, where for moderately controversial topics, attitude change was higher than for strongly controversial topics (\textbf{RQ3}). 
Participants began with more favorable attitudes toward moderately controversial topics, but early change (\textit{i.e.,} in $\Delta_1$) was smaller for these topics relative to highly controversial ones---consistent with ceiling effects, early consolidation, or lower initial uncertainty  (\textbf{RQ3}). In the second segment ($\Delta_2$), controversiality did not shape attitude change.
This finding is not surprising, given that opinions on topics with varying controversiality are often grounded in core values, and thereby, be more or less resistant to change~\cite{eaton2008attitude}.

%With further analysis of user factors, we did not observe any moderating effects of AI literacy, intellectual humility, need for cognition, and user engagement (\textbf{RQ4}). 

\paratitle{Understanding user preferences and trade-offs}. 
We asked to indicate their preferences among the information mediums to engage with: SERPs, AI-generated podcasts, or a combination of both. Our results revealed that a majority of participants (58\%) preferred the combination of both mediums in sequence, while fewer favored either only podcasts (24\%) or SERPs (19\%) for learning the topic. The rationales provided for these preferences, as indicated by the participants, are summarized in Table~\ref{tab-reasons}. 
In their open-ended feedback, participants highlighted a core tension in their information-seeking experiences: a trade-off between \textit{depth} and \textit{accessibility}. Podcasts were recognized for being engaging, memorable, and easy to incorporate into daily routines. However, participants also acknowledged limitations of podcasts, particularly the difficulty of verifying sources and the potential lack of depth or breadth when compared to text-based formats. In contrast, SERPs were valued for offering comprehensive coverage and clear source traceability, though participants noted that this came at the cost of increased cognitive effort and time.

The complementary nature of these strengths and weaknesses suggests that users may benefit most from a hybrid information access ecosystem, that enables smooth transitions between the narrative richness of podcasts and the structured, traceable detail of search results. Participants reflected that such an approach can help them quickly grasp key ideas while still allowing them to explore deeper details when needed.
However, participants also noted that this hybrid model introduces new challenges; notably, increased time demands and the effort required to navigate between modalities. These observations raise important design considerations for future information systems: \textit{How might search engines and content platforms integrate audio narratives and textual results in ways that reduce friction rather than amplify it}?

We identify several promising directions for future work. First, future systems might embed AI‑generated podcast snippets directly within SERPs, giving users a rapid auditory overview alongside traditional links. Second, drawing on advances in retrieval‑augmented generation~\cite{fan2024survey}, AI‑produced podcasts could be paired with automatically retrieved evidence, providing the dual benefits of engaging audio guidance and text-based verifiability. A practical example already emerging in some platforms is the inclusion of linked sources within podcast transcripts, enabling users to move seamlessly from listening to exploring relevant SERPs on demand. Together, these approaches highlight opportunities to design hybrid information experiences that preserve accessibility while enhancing depth and trustworthiness.

\paratitle{Caveats and Limitations}. Given the complexity of studying how information medium sequences (SERPs, podcasts) influence user attitude changes, there are a few important limitations of our work. It is worth noting that we leveraged NotebookLM to create the AI-generated podcasts, which result in two speakers (female, male) discussing the content. Understanding how the observed effects would change based on such configurations of podcasts is beyond the scope of our work, but it is an important avenue for future research.   In our study, we considered users being exposed to consistent viewpoint biases across the two information mediums that they encountered in sequence. Future work should explore the effect of medium sequence on attitude change, varying viewpoint biases across the mediums encountered in sequence.

%% file: 6_Conclusions.tex
\section{Conclusions}

Our study is the first of its kind and addresses an important research gap by offering a systematic analysis of how information modality, content bias, and user traits interact to shape and shift user attitudes. Through a controlled study, we explored how information medium sequences (with SERPs, Podcasts) influence user attitude change, and how such effects are influenced by viewpoint biases in the content that is consumed, and by the degree of controversiality of the topics. Our findings present strong evidence that calls for thoughtful design and integration of different information media in search ecosystems. Given the growing consumption of information across different mediums and modalities and the evolving information-seeking behavior today, we believe that the CHIIR community is uniquely positioned to help navigate these complex and important considerations. We hope that our work inspires further research on responsible opinion formation in multimodal contexts.